# Radiative Recoil in the Vavilov-Cherenkov Effect with Excitation of Surface Plasmons

V.S.Zuev* and G.Ya.Zueva**

*The P.N.Lebedev Physical Institute of RAS*
*119991 Leninsky pr., 53, Moscow*
**The A.M.Prokhorov General Physics Institute of RAS*
*119991 Vavilov str., 38, Moscow*
vizuev@sci.lebedev.ru

Summary
An effect that is an analogue to the Vavilov-Cherenkov effect in a uniform space would be observed in a space with a boundary of well reflecting light metal-dielectric. The main difference is in that quanta of some electron volt would be excited by an electron of energy of some tens of electron volts. This would become possible because the surface plasmon velocity is very small as compared with the velocity of waves of a uniform space. Beetwen excited plasmons there is a plasmon with a linear momentum as high as the doubled initial momentum of the electron.

# Реакция излучения в эффекте Вавилова-Черенкова с возбуждением поверхностных плазмонов

В.С.Зуев, Г.Я.Зуева

Физический ин-т им. П.Н.Лебедева РАН
Институт общей физики им. А.М.Прохорова РАН
vizuev@sci.lebedev.ru

Аннотация
Эффект, аналогичный эффекту Вавилова-Черенкова в однородном пространстве, наблюдается в пространстве с границей раздела хорошо отражающий свет металл-диэлектрик. Отличие заключается в том, что кванты с энергией в несколько электрон вольт возбуждает электрон с энергией в десятки электрон вольт. Это возможно в силу того, что скорость поверхностных плазмонов мала в сравнении со скоростью волн однородного пространства. Среди возбуждаемых плазмонов есть плазмон с импульсом, равным удвоенному начальному импульсу электрона.

# Реакция излучения в эффекте Вавилова-Черенкова с возбуждением поверхностных плазмонов

В.С.Зуев, Г.Я.Зуева

Физический ин-т им. П.Н.Лебедева РАН
Институт общей физики им. А.М.Прохорова РАН
vizuev@sci.lebedev.ru

Электрон, движущийся в пространстве со слоем металла, будет возбуждать поверхностные плазмоны на границе раздела металл-диэлектрик /1/. Будут возбуждаться плазмоны с фазовой скоростью, которая меньше скорости электрона. Явление сходно с явлением Вавилова-Черенкова /2/.

В однородном диэлектрике черенковское излучение возбуждают электроны со скоростью, по порядку величины равной, хотя и меньше, скорости света в вакууме. Скорости поверхностных плазмонов меньше скорости света в вакууме во многие десятки и сотни раз. Это означает, что в такое же число раз может быть меньше скорость электрона, возбуждающего плазмон.

Энергия фотонов оптических частот в возбуждаемых плазмонах составляет приблизительно $3\,eV$. Процесс с испусканием таких фотонов сопровождается эффектом отдачи, что проявляется в уменьшении энергии электрона и в изменении его импульса. Расчет эффекта отдачи составляет содержание данной статьи.

Рассмотрение эффекта отдачи проведем так, как это сделано в /3/. Отличия заключаются в рассмотрении поверхностных волн.

Будем исходить из законов сохранения энергии и импульса. Будем рассматривать нерелятивистский электрон $v_e/c \ll 1$.

$$\frac{m_e v_e^2}{2} = \frac{m_e v_{e'}^2}{2} + \hbar\omega', \tag{1}$$

$$m_e \vec{v}_e = m_e \vec{v}_{e'} + \hbar \vec{k}'. \tag{2}$$

Здесь $m_e$, $\vec{v}_e$ и $\vec{v}_{e'}$ - масса электрона, его начальная и конечная скорости, $\omega'$ и $\vec{k}'$ - частота и волновой вектор испущенного фотона. Волновой вектор $\vec{k}'$ испущенного фотона и первоначальная скорость $\vec{v}_e$ располагаются под углом $\Omega$ друг к другу, $\cos\Omega = \vec{k}'\cdot\vec{v}_e / k' v_e$. Из (1), (2) получаем

$$k'^2 - 2k' m_e v_e \cos\Omega/\hbar + 2m_e\omega'/\hbar = 0, \tag{3}$$

$$k' = \frac{m_e v_e}{\hbar}\cos\Omega\left[1 \pm \sqrt{1 - \frac{2\hbar\omega'}{m_e v_e^2 \cos^2\Omega}}\right]. \tag{4}$$

При $2\hbar\omega'/m_e v_e^2 \cos^2\Omega \ll 1$ импульс фотона равен либо

$$\hbar k' \approx 2 m_e v_e \cos\Omega, \tag{5a}$$

либо

$$\hbar k' \approx \hbar\omega'/v_e \cos\Omega \tag{5b}$$

в зависимости от того, какой знак мы выбираем в формуле (4).

Начнем рассмотрение со случая, описываемого формулой (5a), причем при $\cos\Omega = 1$. На рис.1 показаны дисперсионные кривые для симметричного (кривая 1) и антисимметричного (кривая 2) плазмонов на пленке $Au$ в вакууме. Толщина пленки $2\,nm$, плазменная частота $\omega_{pl} = 6.75\cdot 10^{15}\,rad/s$ (прямая 3 - $\omega_{pl}/\sqrt{2}$). Прямая 4 – функция $kv_e$. Пересечение прямой 4 с какой-либо из дисперсионных кривых 1 или 2 дает волновое число $k'$ и частоту $\omega'$ для соответствующего плазмона.

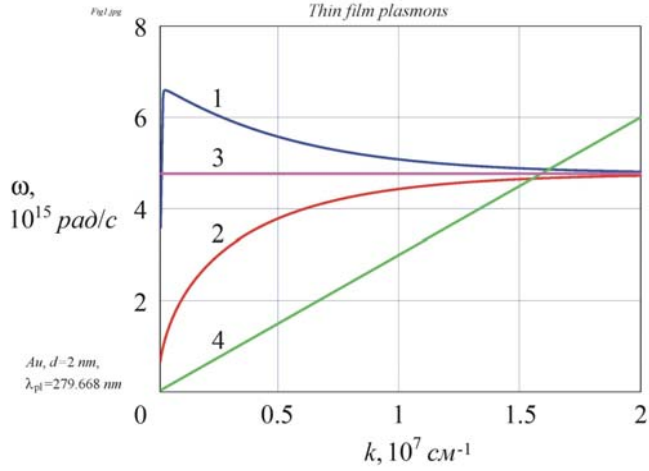 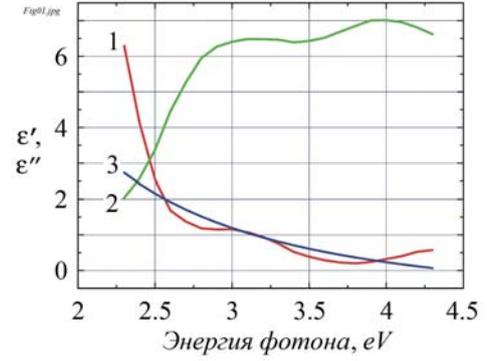

Рис.1.                                                        Рис.2.

Сравнение данных о свойствах $Au$ из /4/ с данными из /5/ показывает, что точка $\varepsilon = -2$ совпадает у этих авторов. Поведение кривых несколько различается, но мало. Проницаемость оказывается равной $\varepsilon = -2$ при $2.565\ eV$. Для описания проницаемости примем формулу Друде $\varepsilon = 1 - \omega_{pl}^2/\omega^2$. Тогда $\varepsilon = -2$ при $\hbar\omega_{pl} = 4.445\ eV$ и $\hbar\omega = 2.565\ eV$.

На рис.2. изображены действительная часть, $\varepsilon'$, кривая 1, и мнимая часть, $\varepsilon''$, кривая 2, проницаемости $Au$ по данным из /4/ и аппроксимирующая для $\varepsilon'$ кривая 3 по формуле Друде. В области $2.5 \div 4.0\ eV$ диэлектрическая проницаемость золота хорошо описывается формулой Друде с $\hbar\omega_{pl} = 4.445\ eV$.

При большом импульсе фотона $\hbar k'$ частота испускаемого фотона $\omega'$ мало отличается от $\omega_{pl}/\sqrt{2}$. С помощью (2) получаем, что импульс рассеянного электрона равен

$$m_e v_{e'} = m_e v_e - 2 m_e v_e = -m_e v_e. \quad (15)$$

Рассеянный электрон летит в обратную сторону по отношению к первоначальному направлению с импульсом, по величине равным первоначальному.

Рассмотрим численный пример. Возьмем массу в 10 раз меньше электронной: $m_e^* = 0.1 \cdot m_e = 9 \cdot 10^{-29}\ g$. Такой может быть эффективная масса электрона в веществе. Если брать массу больше, то возникают не реализуемые цифры, в частности, слишком малая скорость плазмона. Энергия фотона

$$\hbar\omega' \approx \hbar\omega_{pl}/\sqrt{2} = 4.445\ eV/\sqrt{2} = 3.15\ eV.$$

Будем считать, что

$$m_e^* v_e^{*2} = 31.5\ eV \cdot 1.6 \cdot 10^{-12}\ erg/eV = 5.04 \cdot 10^{-11}\ erg.$$

При таком выборе условие $2\hbar\omega'/m_e^* v_e^2 \cos^2\Omega \ll 1$ хорошо соблюдается. Скорость электрона при этом оказывается равной $v_e^* = 5.3 \cdot 10^8\ cm/s$, а импульс фотона -

$\hbar k' = 2 m_e^* v_e^* = 0.954 \cdot 10^{-19}\ g \cdot cm/s$. Волновое число при таком импульсе равно $k' = 0.9 \cdot 10^8\ cm^{-1}$, длина волны плазмона $\lambda' \approx 7 \cdot 10^{-8}\ cm$. Для фотона с энергией $3.15\ eV$ длина волны в вакууме равна $280\ nm$. Длина волны возбуждаемого плазмона в 400 раз меньше этой величины. Скорость плазмона равна $v' = 7.5 \cdot 10^7\ cm/s$.

По формуле (6b) получаем фотоны в направлении вперед со скоростью, близкой хотя и меньше скорости $v_e$. Энергия электрона уменьшается на величину $\hbar\omega' \approx \hbar\omega_{pl}/\sqrt{2}$.

Подведем итог. Эффект, аналогичный эффекту Вавилова-Черенкова в однородном пространстве, наблюдается также в пространстве с границей раздела хорошо отражающий свет металл-диэлектрик. Отличие заключается в том, что кванты с энергией в несколько электрон вольт возбуждает электрон с энергией в десятки электрон вольт. Это возможно в силу того, что скорость поверхностных плазмонов мала в сравнении со скоростью волн однородного пространства. Среди возбуждаемых плазмонов есть плазмон с импульсом, равным удвоенному начальному импульсу электрона.

1. В.С.Зуев, Г.Я.Зуева. Оптика и спектроскопия, т.108, 680-683 (2010)
2. П.А.Черенков. Труды ФИАН, т.2, вып 4, стр 3-62 (1944)
3. В.Л.Гинзбург. УФН, т.69, 537-564 (1959)
4. E.D.Palik. Handbook of Optical Constants of Solids. Academic Press, San Diego, 1998
5. P.B.Johnson, R.W.Christy. Phys. Rev. B, v.6, 4370-4379 (1972)